\documentclass[12pt]{article}
\usepackage{epsfig}

\def\beq{\begin{equation}}
\def\eeq#1{\label{#1}\end{equation}}
\def\eeqn{\end{equation}}

\def\beqa{\begin{eqnarray}}
\def\eeqa#1{\label{#1}\end{eqnarray}}
\def\eeqan{\end{eqnarray}}

\let\bar=\overbar

\def\Dslash{\not{\hbox{\kern-4pt $D$}}}
\def\dslash{\not{\hbox{\kern-2pt $\del$}}}

\def\msb{{\bar{\ssstyle M \kern -1pt S}}}

\def\Title#1{\begin{center} {\Large {\bf #1} } \end{center}}

\begin{document}

\Title{Charm and beauty structure of the proton}

\begin{center}{\large \bf Contribution to the proceedings of HQL06,\\
Munich, October 16th-20th 2006}\end{center}

\bigskip\bigskip

\begin{raggedright}  

{\it Riccardo Brugnera\index{Brugnera, R.}\footnote{On behalf of 
the H1 and ZEUS Collaborations.}\\
Dipartimento di Fisica dell'Universit\'{a} di Padova and\\
INFN Sezione di Padova\\
E-mail:riccardo.brugnera@pd.infn.it}
\bigskip\bigskip
\end{raggedright}

\section{Introduction}
\label{sec:Intro}
Heavy quarks production is an important testing ground for
quantum chromodynamics (QCD), because QCD calculations
are expected to be reliable if a hard scale is present
in the process. In heavy quarks production a hard scale
is provided by the quark mass.
Moreover heavy quarks production can give direct access
to the gluon density in the proton due to the fact
that it proceeds, in QCD, almost exclusively via
photon-gluon fusion, where a photon from the
incoming electron interacts with a gluon in the proton
giving an heavy quark-anti-quark pair.
Results will be shown both for deep-inelastic
scattering (DIS), where the virtuality of the exchanged
boson $Q^2$ is large, and photo-production, where the
$Q^2$ is equal to zero.
Various experimental techniques are used in order to select
charm and beauty events, ranging from the measurement of
$D^{*}$ cross section to impact parameter analyses.
The results are found to be compatible with the predictions of
perturbative QCD (pQCD).

This paper is organized as follows. The relevant features of the HERA 
collider and of the H1 and ZEUS detectors are described in section  
\ref{sec:HERA}. In section \ref{sec:Theory}, an introduction to the 
physics of heavy quarks production in $ep$ collisions is given. 
The sections \ref{sec:Charm-tagging} and \ref{sec:Charm-results} 
illustrate the tagging methods and the experimental results for the 
charm quark, while \ref{sec:Beauty-tagging} and \ref{sec:Beauty-results} 
do the same for the beauty quark. The charm and beauty structure functions  
are presented in section \ref{sec:F2}. The results obtained for the gluon 
polarization by the COMPASS Collaboration are described in section 
\ref{sec:COMPASS}. Finally the conclusions are drawn in section 
\ref{sec:conclusions}.

\section{The HERA collider and its two multipurpose experiments: H1 and ZEUS}
\label{sec:HERA}
\begin{figure}[htb]
\begin{center}
\epsfig{file=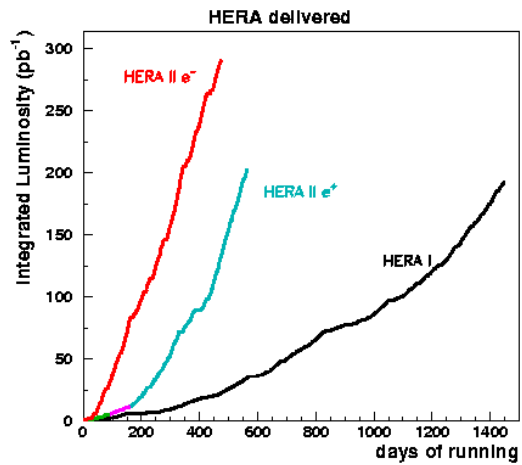,height=2.5in}
\caption{The integrated luminosity delivered by HERA, subdivided into HERA-I 
period and HERA-II, versus the days of running. The HERA-II period is 
furthermore divided in HERA-II with electrons (HERA-II $e^{-}$) and HERA-II 
with positrons (HERA-II $e^{+}$).}
\label{fig:HERALumi}
\end{center}
\end{figure}

HERA is the first $ep$ collider and consists of two separate rings of circumference 
6.3 km, one a warm magnet electron (or positron) ring with maximum energy 30 GeV 
and the other a superconducting magnet proton ring of maximum energy 920 GeV. 
The rings are brought together at four intersection regions, two of them are 
occupied by the experiments H1 and ZEUS. 
The HERA life can be divided in two parts: HERA-I from 1992 to 2000 and 
HERA-II from 2003 to the middle of 2007. 
During the first period, HERA worked with $e^{\pm}$ beam of 27.5 GeV while 
the energy of the proton beam was raised from 820 GeV to 920 GeV. 
The beam spot had the dimension of 150$\times$30 $\mu$m$^2$ and the 
integrated luminosity collected by each experiments was about 130 pb$^{-1}$.
At the end of 2000 there was a long shutdown, in which both HERA and the 
two experiments made important upgrades. In 2003 HERA started its 
functioning and it is to foreseen to work up to the middle of 2007. 
During the HERA-II period the energies of the lepton and proton beams  
remained unchanged: 27.5 GeV and 920 GeV respectively. A reduced beam 
spot (80$\times$20 $\mu$m$^2$) and more reliable beams operation 
have enhanced a lot the delivered luminosity to the experiments  
($\sim$ 180 pb$^{-1}$ per experiment from 2003 to 2005). In Fig. 
\ref{fig:HERALumi} the integrated luminosities 
per period are shown as function of the day of the run.

\begin{figure}[htb]
\begin{center}
\includegraphics[width=0.7\textwidth]{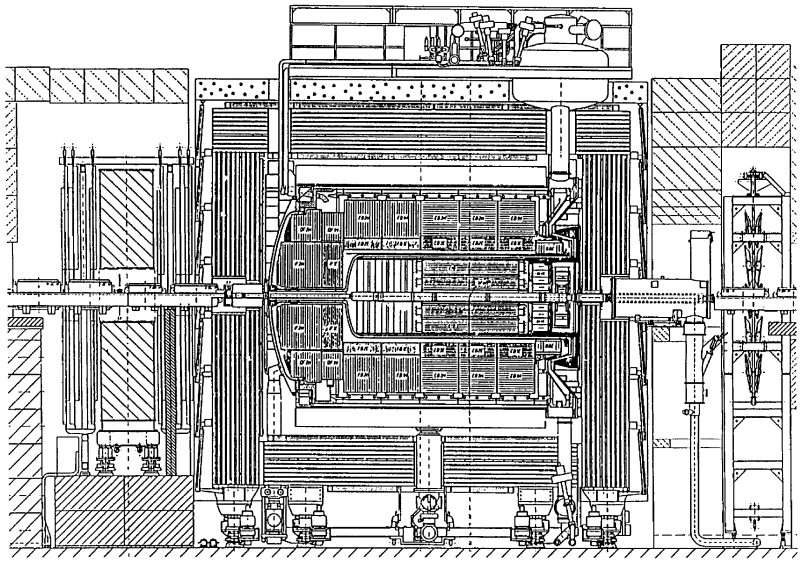}
\caption{Cross sectional view of the H1 detector.}
\label{fig:H1view}
\end{center}
\end{figure}

\begin{figure}[htb]
\begin{center}
\includegraphics[width=0.9\textwidth]{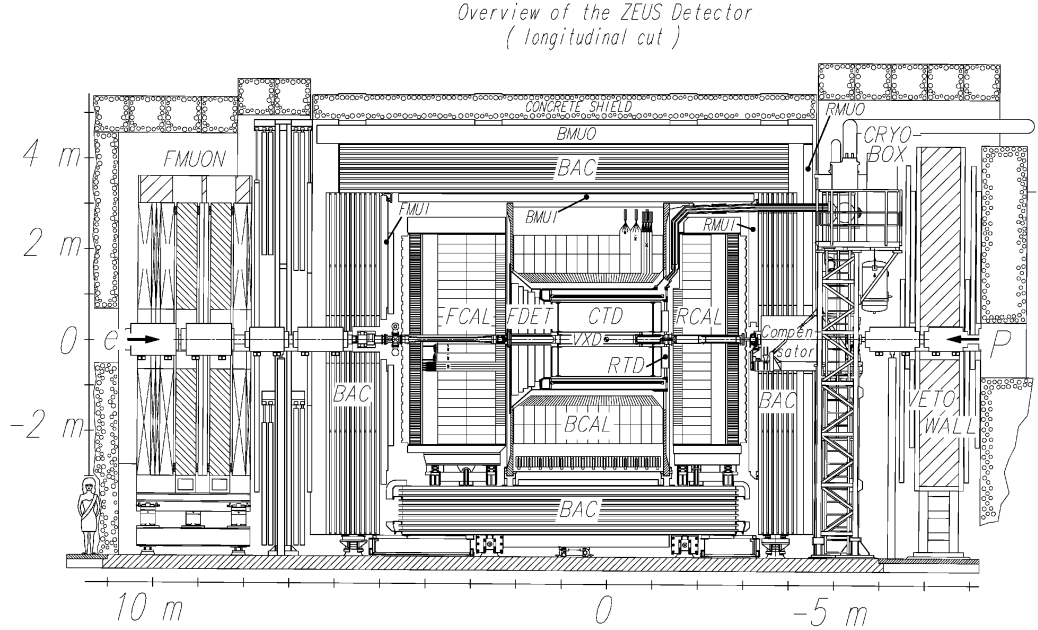}
\caption{Cross sectional view of the ZEUS detector.}
\label{fig:ZEUSview}
\end{center}
\end{figure}

The H1 \cite{H1} and ZEUS \cite{ZEUS} 
detectors are general purpose detectors with nearly hermetic 
calorimetric coverage. They are designed in order to investigate all aspects of high energy $ep$ collisions. In particular both the scattered electron and the 
hadronic system in a hard $ep$ interaction are measured. They are differentiated principally by the choices made for the calorimetry. The H1 collaboration has 
stressed electron identification and energy resolution, while the ZEUS 
Collaboration has put its emphasis on optimizing calorimetry for 
the hadronic measurements. 
The detector designs reflect these different emphases. The H1 detector has a 
large diameter magnet encompassing the main liquid argon calorimeter, while 
the ZEUS detector has chosen a uranium-scintillator sampling calorimeter with 
equal response to electrons and hadrons. 
Cross sectional view of the H1 and ZEUS detectors are presented in Fig. 
\ref{fig:H1view} and \ref{fig:ZEUSview} respectively.

\section{Production of Heavy Quarks in $ep$ collisions}
\label{sec:Theory}
\begin{figure}[htb]
\begin{center}
\epsfig{file=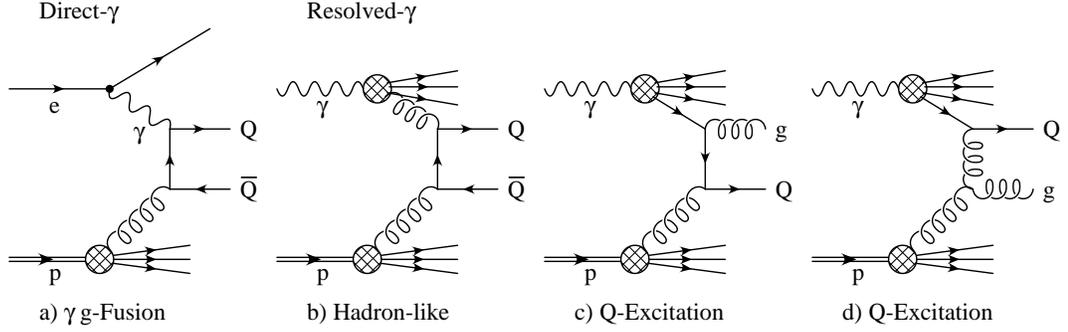,height=1.8in}
\caption{Heavy Quarks production processes in leading order pQCD. }
\label{fig:feynman}
\end{center}
\end{figure}

In pQCD, at leading order (LO), two distinct classes of processes contribute 
to the production of heavy quarks (charm and beauty) in $ep$ 
collisions at HERA. 
In direct-photon processes (Fig. \ref{fig:feynman}a), 
the photon emitted from the electron 
enters the hard process $\gamma g \rightarrow Q \overline{Q}$ directly. 
In resolved-photon processes (Fig. \ref{fig:feynman}b to \ref{fig:feynman}d), 
the photon fluctuates into a hadronic state before the hard interaction and 
acts as a source of partons, one of which takes part in the hard interaction. 
Resolved photon processes are expected to contribute significantly in the 
photo-production regime, in which the photon is quasi-real, and to be 
suppressed towards higher $Q^2$. 

Next-to-leading order (NLO) calculations in several schemes are available 
\cite{NLO-DIS,NLO-PHP}.
In DIS regime, all approaches assume that $Q^2$ and heavy quark mass $m_Q$ provide a hard enough scale to allow the applicability of pQCD and to garantee the validity of the factorization theorem. In photo-production 
regime the hard scale is
given by the transverse momentum of the heavy quark $p_{t,Q}$ and $m_Q$.
In the fixed-order, or ``massive'', scheme
\footnote{The scheme is often referred to as fixed flavour number scheme 
(FFNS).}, $u,d,s$ are the only active 
flavours in the structure functions of the proton and photon. The heavy quarks are assumed to be produced only at perturbative level via photon-gluon fusion. 
This scheme is expected to work well in regions where $p_{t,Q}^2 \sim m_{Q}^2$ 
(if in photo-production regime) or where $Q^2 \sim  m_{Q}^2$ (if in DIS regime).
At higher transverse momenta or $Q^2$, calculations based on this scheme can break down due to large logarithms $\sim ln (p_{t,Q}^2/m_{Q}^2)$ ($\sim ln (Q^2/m_{Q}^2)$). In this case the resummed, or ``massless'', scheme
\footnote{The scheme is often referred as the zero mass variable flavour 
number scheme (ZMVFNS).} \cite{NLO-massless} should be applicable. 
In this scheme, charm and beauty are regarded as active flavours (massless 
partons) in the structure functions of the proton and photon and are fragmented from massless partons into massive hadrons after the hard process.   
There are also calculations\footnote{The scheme is commonly referred to as 
variable flavour number scheme (VFNS).}  
which tempt to treat the heavy quarks correctly 
for all $Q^2$. Therefore, at low $Q^2$, an heavy quark is produced dynamically 
through the boson-gluon fusion process, 
whereas, at high $Q^2$, heavy quark parton densities 
are introduced. The transition between the two extremes is treated in 
different way by different authors \cite{NLO-mixed}.

\section{Charm production: tagging methods}
\label{sec:Charm-tagging}
\begin{figure}[htb]
\begin{center}
\includegraphics[width=0.5\textwidth]{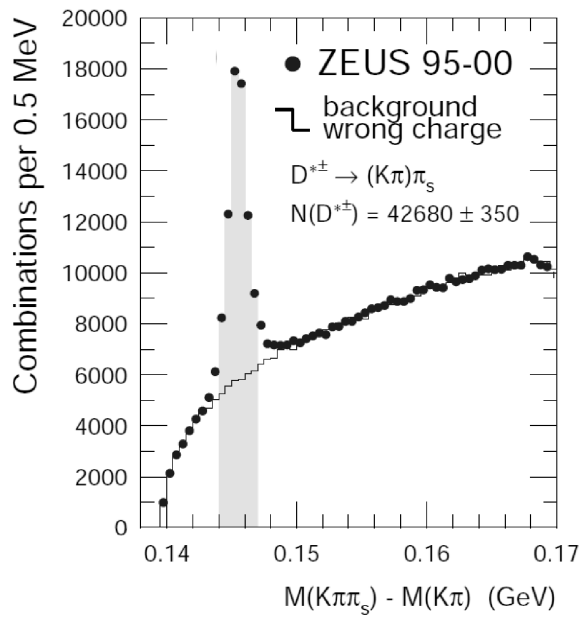}
\caption{
The distribution of the mass difference, 
$ \Delta M $ = $M(K\pi \pi_s) - M(K\pi)$, 
for $D^{*}$ candidates. The $D^{*\pm}$ candidates (dots) 
are shown compared to the 
wrong charge combinations (histogram). 
The shaded region shows the signal region. The number of $D^{*}$
mesons is determined by subtracting the wrong charge background.}
\label{fig:DstarTag}
\end{center}
\end{figure}

The main method used for charm tagging is the identification of the $D^{*}$ mesons using the decay channel $D^{*+} \rightarrow D^{0} \pi^{+}_{s}$ with the 
subsequent decay $D^{0} \rightarrow K^{-} \pi^{+}$, where $\pi_s$ refers to 
the low momentum $\pi$ in the decay. The decay particles of the $D^{*}$ meson are reconstructed in the central detector, usually without particle 
identification. In Fig. \ref{fig:DstarTag} it is shown a distribution 
of the mass difference $\Delta M = M(K\pi \pi_s) - M(K\pi)$ from the  
ZEUS Collaboration. 
A clear signal is seen around the nominal value 
$M(D^{*})-M(D^{0})$. In order to mantain under control the combinatorial 
background, various cuts are made on the $p_t$ of the tracks and on the energy of the event. 
Of course also other charmed hadrons were identified and analyzed, such as 
$D^{+}, D_s, \Lambda_c$, but with less statistics. 
Finally, the sistematic use of the vertex detectors, first implemented in H1 
and now also in ZEUS, is changing dramatically the perspective of the 
physical analysis in the charm sector as it already happened in the beauty 
one (see section \ref{sec:Beauty-tagging} and \ref{sec:Beauty-results}).

\section{Charm production: experimental results}
\label{sec:Charm-results}
\begin{figure}[htb]
\begin{center}
\includegraphics[width=0.40\textwidth]{fig4d_sup_h1.epsi}
\includegraphics[width=0.55\textwidth]{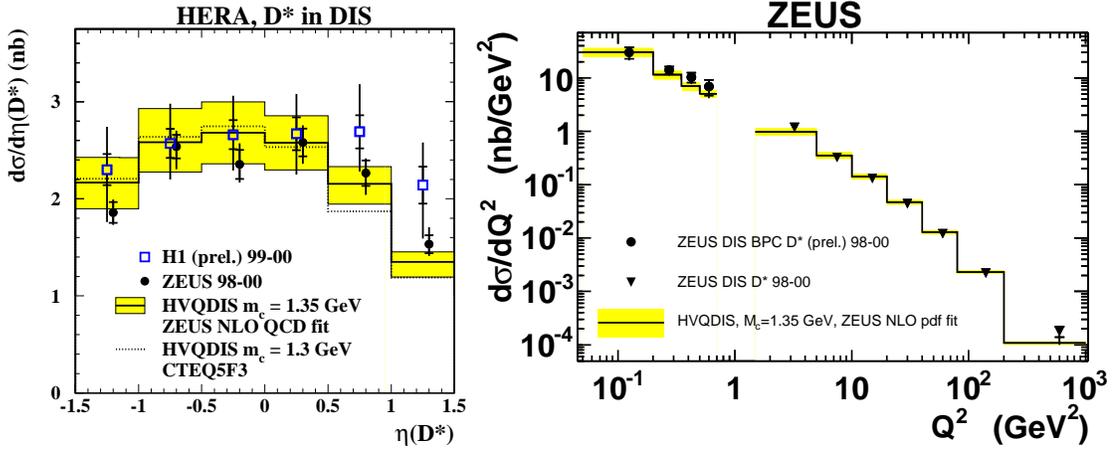}
\caption{On the left, differential $D^{*}$ cross sections as a function of $\eta(D^{*}$), filled points from the ZEUS experiment and empty squares from the H1 
experiment. The bands  
are the NLO predictions of HVQDIS.
On the right, differential $D^{*}$ cross sections as a function of $Q^2$ for low $Q^2$ (dots) 
and from results on $D^{*}$ production in DIS (triangle) compared to the NLO predictions 
from HVQDIS. The data come from the ZEUS Collaboration.}
\label{fig:DstarEta}
\end{center}
\end{figure}

The status of the charm analysis can be summarized by the two plots of 
Fig. \ref{fig:DstarEta}, where the differential $D^{*}$ cross section 
as a function of the pseudo-rapidity\footnote{The pseudo-rapidity $\eta$ 
corresponding to a polar angle $\theta$ (
measured respect to the positive z-axis, corresponding to the the incoming proton 
beam direction) is given by $\eta$ = -ln tan($\theta/2$).} of the $D^{*}$ mesons,
$\eta(D^{*}$) on the left, and the differential $D^{*}$ cross sections as 
a function of $Q^2$ on the right are shown \cite{DstarData}. 

\begin{figure}[htb]
\begin{center}
\epsfig{file=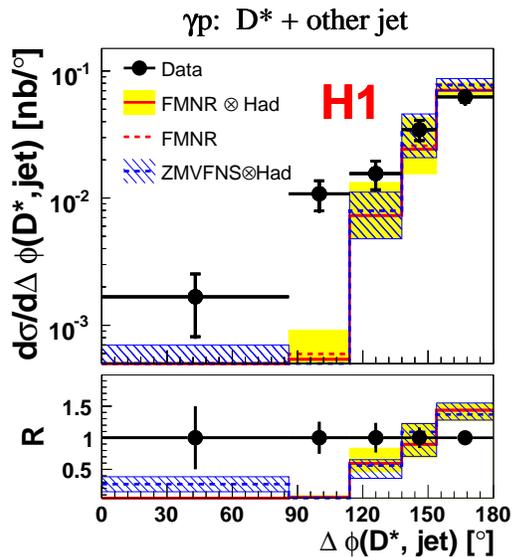,height=3.0in}
\caption{$D^{*}+$jet cross sections as function of $\Delta \phi$($D^{*},$jet) compared 
with the predictions of the NLO calculations FMNR and ZMVFNS. 
}
\label{fig:DstarJet}
\end{center}
\end{figure}

\begin{figure}[!htb]
\begin{center}
\epsfig{file=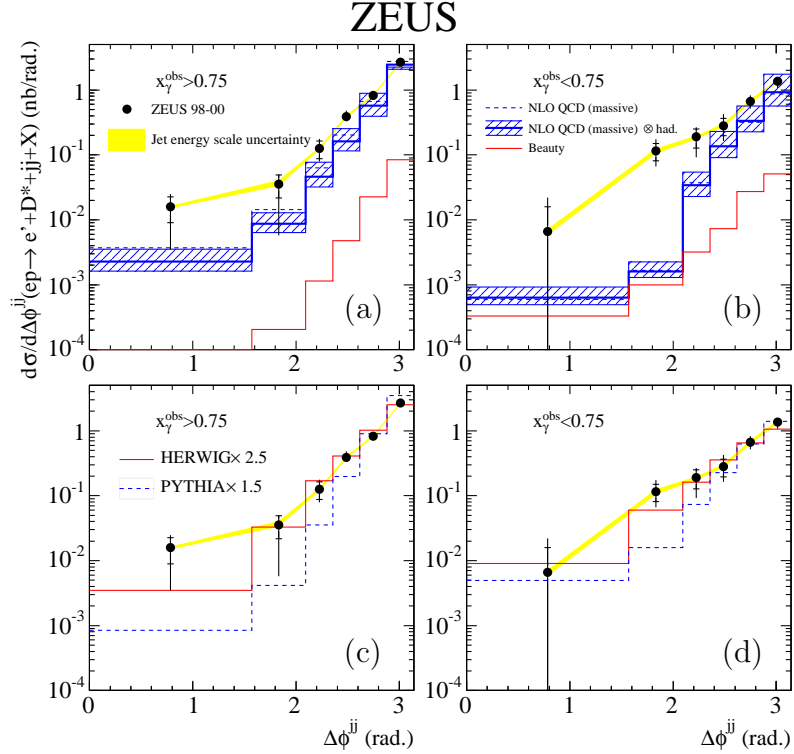,height=4.0in}
\put(-175.0,165.0){(a)}
\put(-32.0,165.0){(b)}
\put(-175.0,35.0){(c)}
\put(-32.0,35.0){(d)}
\caption{Cross section for the process $ep \rightarrow e+D^{*}+jj+X$ 
separated into (a,c) direct enriched ($x_{\gamma}^{obs} > 0.75$) 
and (b,d) resolved enriched  ($x_{\gamma}^{obs} < 0.75$). The data (solide dots) are compared (a,b) to the massive QCD prediction with (solid line) and without (dotted line) hadronisation corrections applied. The theoretical uncertainties (hatched band) come from the change in scales simultaneously with the change in charm mass. The beauty component is also shown (lower 
histogram). The data are also compared (c,d) with HERWIG (solid line) and PYTHIA (dashed line) MC predictions multiplied by the indicated factors. The data come from the ZEUS Collaboration.
}
\label{fig:DstarDijets}
\end{center}
\end{figure}

The plot on the left of Fig. \ref{fig:DstarEta} shows the good agreement 
between the ZEUS and H1 data. The bands in both plots represent the NLO predictions 
using the HVQDIS program \cite{HVQDIS}, the widths of the bands correspond to the 
uncertainties in the mass of the charm, in the renormalization and factorization 
scales, in the proton parton density functions and in the fragmentation. 
Rather remarkable is the fact that the $d \sigma /d Q^2$ data are well 
described by NLO calculations over five orders of magnitude.  
Some discrepancies between data and theory are seen 
in photoproduction: $D^{*}$ 
photoproduction cross sections \cite{DstarPhoto} as function of 
the transverse momentum, 
$p_T(D^{*})$, and $\eta(D^{*})$ show that the predictions from NLO QCD are too low for 
$p_T(D^{*}) > $ 3 GeV and $\eta(D^{*}) >$ 0. Part of this deficit may be due to hadronisation 
effects. The predictions for single jet and dijet production accompanied by 
a $D^{*}$ meson should have smaller uncertainties from these effects. 
For that aim the following correlations were studied: the 
difference in the azimuthal angle, $\Delta \phi$($D^{*},$jet), between 
the $D^{*}$ and a jet not containing the $D^{*}$ meson and those between 
the two jets of highest transverse energy,  $\Delta \phi^{jj}$, and 
the squared transverse momentum of the dijet system, $(p_{T}^{jj})^2$. 
For the LO 2 $\rightarrow$ 2 process, the two jets, or the 
$D^{*}$ and a jet not containing the $D^{*}$ meson, are produced back-to-back 
with $\Delta \phi = \pi$ and very low $p_T$. Large deviations from these 
values may come from higher-order QCD effects.
In Fig. \ref{fig:DstarJet} the differential cross section as function of the 
$\Delta \phi$($D^{*},$jet) is shown, a large fraction of the produced 
$D^{*}+$jet combinations deviates 
from back-to-back configuration indicating the importance of higher 
order contributions. The available NLO calculations (massive FMNR \cite{NLO-PHP} 
and ZMVFNS \cite{ZMVFNS}) underestimate significantly the observed 
cross sections in the region $\Delta \phi$($D^{*},$jet)$ < 120^o$.
The cross section $d\sigma/d \Delta \phi^{jj}$, see Fig. \ref{fig:DstarDijets}, 
is reasonable reproduced by the NLO predictions in the direct-enriched region, 
that is $x_{\gamma}^{obs} > 0.75$\footnote{$x_{\gamma}^{obs}$ represents the 
fraction of the photon momentum partecipating to the hard scattering.}
, although the data exhibit a somewhat 
harder distribution. In the resolved-enriched region,  $x_{\gamma}^{obs} < 0.75$, 
the data exhibit a harder spectrum than for $x_{\gamma}^{obs} > 0.75$. 
The NLO prediction of the cross section for $x_{\gamma}^{obs} < 0.75$ has a 
significantly softer distribution compared to the data. The low- $x_{\gamma}^{obs}$ 
region is more sensitive to higher-order topologies not present in the 
massive NLO prediction.  
The predictions from PYTHIA MC \cite{PYTHIA} reproduce neither the shape 
nor the 
normalisation of the data for low and high $x_{\gamma}^{obs}$. However, 
the predictions from the HERWIG MC \cite{HERWIG} give an excellent 
description of the 
shapes of all distributions, although the normalisation is 
underestimated by a factor of 2.5. The fact that a MC programme 
incorporating parton showers can successfully describe the data whereas 
the NLO QCD prediction cannot indicates that the QCD calculation 
requires higher orders. Matching of parton showers with NLO 
calculations such as in the MC@NLO programme \cite{MCNLO}, which 
is not currently available for the processes studied here, should improve the 
description of the data.

\section{Beauty production: tagging methods}
\label{sec:Beauty-tagging}

The H1 and ZEUS Collaborations have presented measurements in which 
the events containing beauty are identified in the following manners:
using high $p_T$ leptons (mainly muons) from semileptonic 
$b$-decays, or using the impact parameters of all tracks coming 
from secondary decay vertices (inclusive lifetime tag analysis), 
or finally using double tagged events ($D^{*}+\mu$, $\mu \mu$). 

In the first method, the transverse momentum $p_{T}^{rel}$ of the 
muon with respect to the axis of the associated jets 
exhibits a much harder spectrum for muons from $b$-decays than
for the other sources. Sometime, in order to enhance the signal to 
noise ratio also the signed impact parameter $\delta$ 
of the muon track with respect to the primary event vertex is used, 
this quantity reflects the lifetime of the particle from which 
the muon decays. The relative contributions from $b, c$ and light 
quarks are determined by a fit to the $p_{T}^{rel}$ distribution or 
to a combined fit to the $p_{T}^{rel}$ and $\delta$ distributions 
using the shapes of Monte Carlo $b, c$ and light quarks distributions 
as templates. 
 
In the second method, the track selection requires full silicon 
vertex detector information. From the measured impact parameter 
$\delta$ a lifetime significance $S = \delta/ \sigma_{\delta}$ is 
calculated. Two independent distributions are constructed. 
$S_1$ is the significance distribution of tracks in events with 
exactly one selected tracks. $S_2$ contains the significances of the 
tracks with the second highest significance for events with 
two or more selected tracks. Events in which the tracks with the 
first and second highest absolute significance have different signs 
are removed from the $S_2$ distribution. 
The subtracted significance distributions are obtained by bin-wise 
subtraction of the numbers of entries on the negative side from 
those on the positive side. The subtraction method substantially 
reduces the systematic uncertainties due to track and vertex 
resolutions. The relative contributions from $b$, $c$ and light 
quarks are determined from a fit to the subtracted $S_1$ and 
$S_2$ distributions and the total number of events, using the shapes 
of Monte Carlo  $b$, $c$ and light quarks distributions as 
templates. 

In the third method, doubled tagged events, events are selected 
containing at least one reconstructed $D^{*}$ and at least one 
muon, $D^{*}+\mu$, or two muons in the final state ($\mu \mu$).
In order to suppress the various types of backgrounds 
the charge and angle correlations of the $D^{*}$ with respect to the muon 
and of the two muons are exploited. These double tagged 
measurements extend to 
significantly lower centre-of mass energies of the $b\bar{b}$ system 
than measurements based on leptons and/or jets with high transverse 
momentum. Furthermore, these double tagged events permit to test 
higher order QCD effects. For instance, in the photon-gluon rest 
frame the angle between the heavy quarks is 180$^o$ at leading 
order, but at NLO it can differ significantly from this value 
due to hard gluon radiation.

\section{Beauty production: experimental results}
\label{sec:Beauty-results}

\begin{figure}[htb]
\begin{center}
\includegraphics[width=0.43\textwidth]{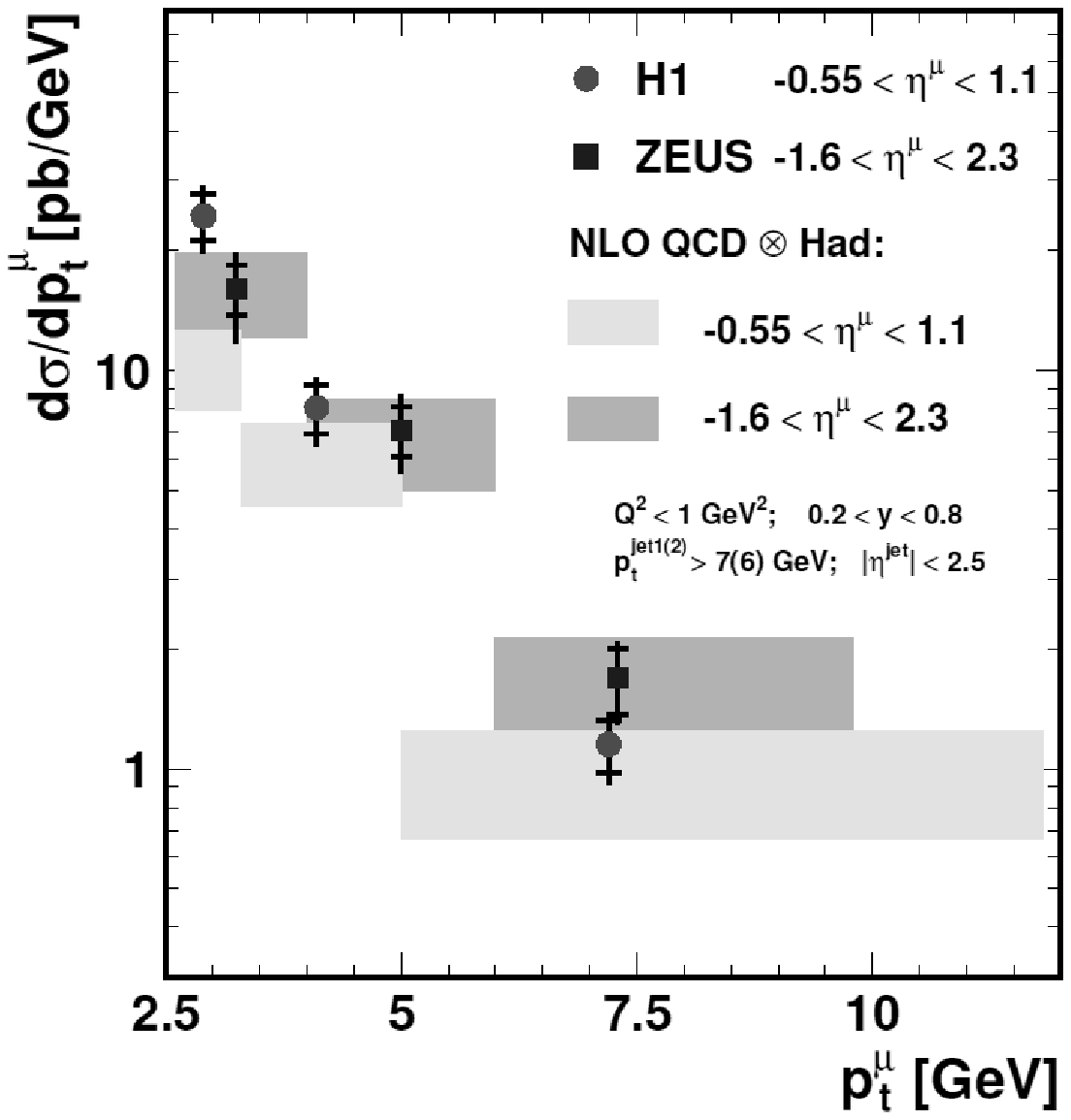}
\includegraphics[width=0.4\textwidth]{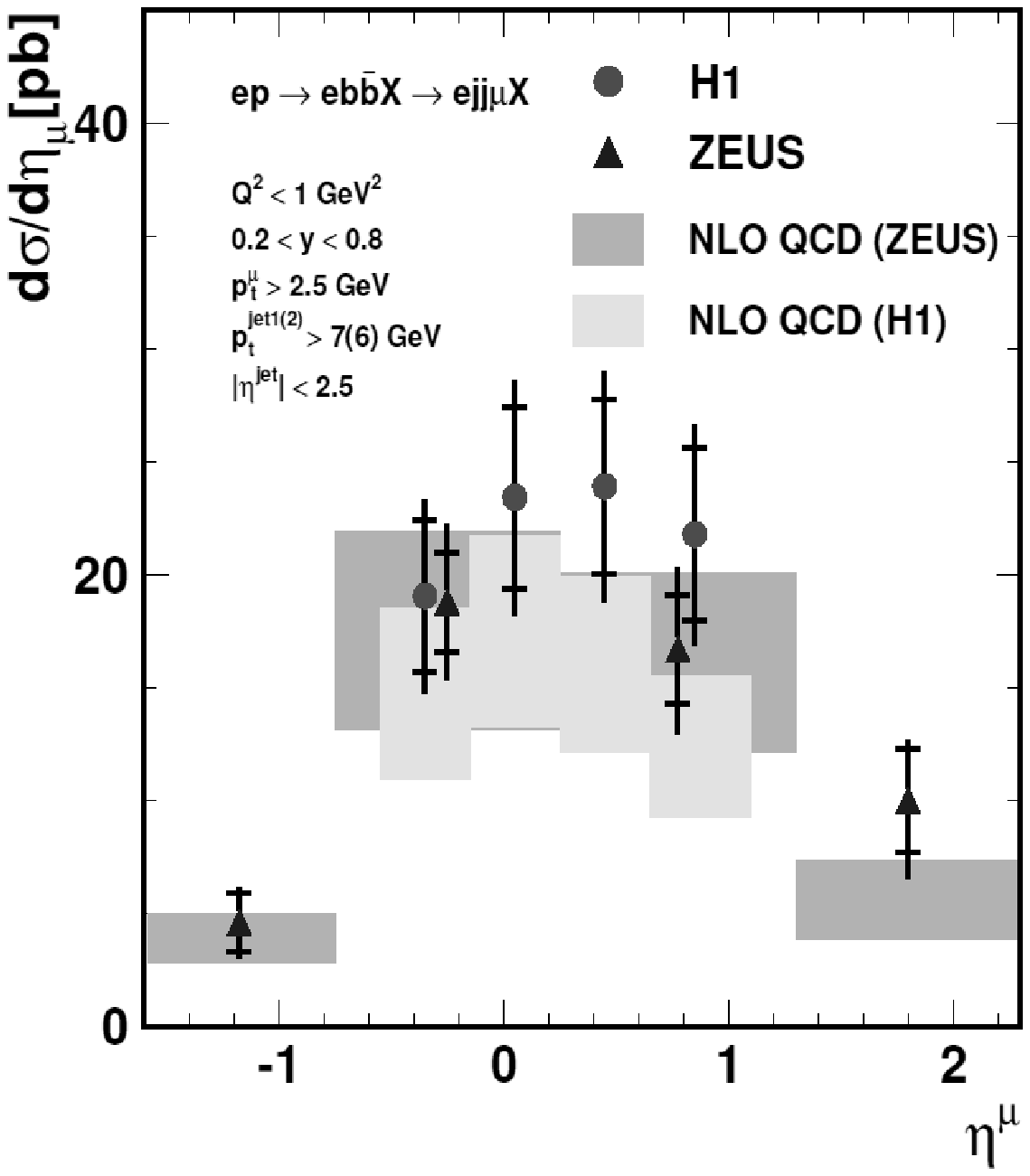}
\caption{Differential cross sections as a function of the muon transverse 
momentum $p_{T}^{\mu}$ (on the left) and the muon 
pseudo-rapidity $\eta^{\mu}$ (on the right), 
for muons coming from $b$ decays in dijet 
events. The two sets of data coming from the H1 and ZEUS experiments were 
measured in different phase space regions. The full error bars are the 
quadratic sum of the statistical (inner part) and systematic 
uncertainties. The bands represent the NLO predictions convoluted 
with their uncertaintites obtained by varying the $b$-quark mass and the 
renormalization and factorization scales.}
\label{fig:Bmudijets}
\end{center}
\end{figure}

Differential measurements from H1 and ZEUS are available for beauty 
production in photoproduction and DIS \cite{BH1mujet},
\cite{BZEUSmujet} using the lepton+jet(s) 
tag method. Figure \ref{fig:Bmudijets} shows the differential 
photoproduction cross sections as a function of the muon transverse 
momentum (on the left) and of the pseudo-rapidity for the process 
$ep \rightarrow eb\bar{b} X \rightarrow e j j \mu$. 
The H1 and ZEUS data, which are in reasonable agreement when they are 
compared in the same phase space region (see the  $d\sigma/d\eta^{\mu}$ 
plot on the right side), are compared to a NLO calculation in the 
massive scheme \cite{NLO-PHP}. The NLO calculations describe the ZEUS data 
well. Comparing with the H1 data, the NLO calculations predict a 
less steep behaviour for the $d\sigma/dp_t^{\mu}$ and is lower 
than the H1 data in the lower momentum bin by roughly a factor of 2.5; 
at higher transverse momenta better agreement is observed. 
In DIS (data not shown), the total cross section measurements made by the 
H1 and ZEUS Collaborations are 
somewhat higher than the predictions. The observed 
excess is pronounced at large muon pseudo-rapidities, low values of $Q^2$ 
and muon transverse momentum. 
        
\begin{figure}[htb]
\begin{center}
\includegraphics[width=0.425\textwidth]{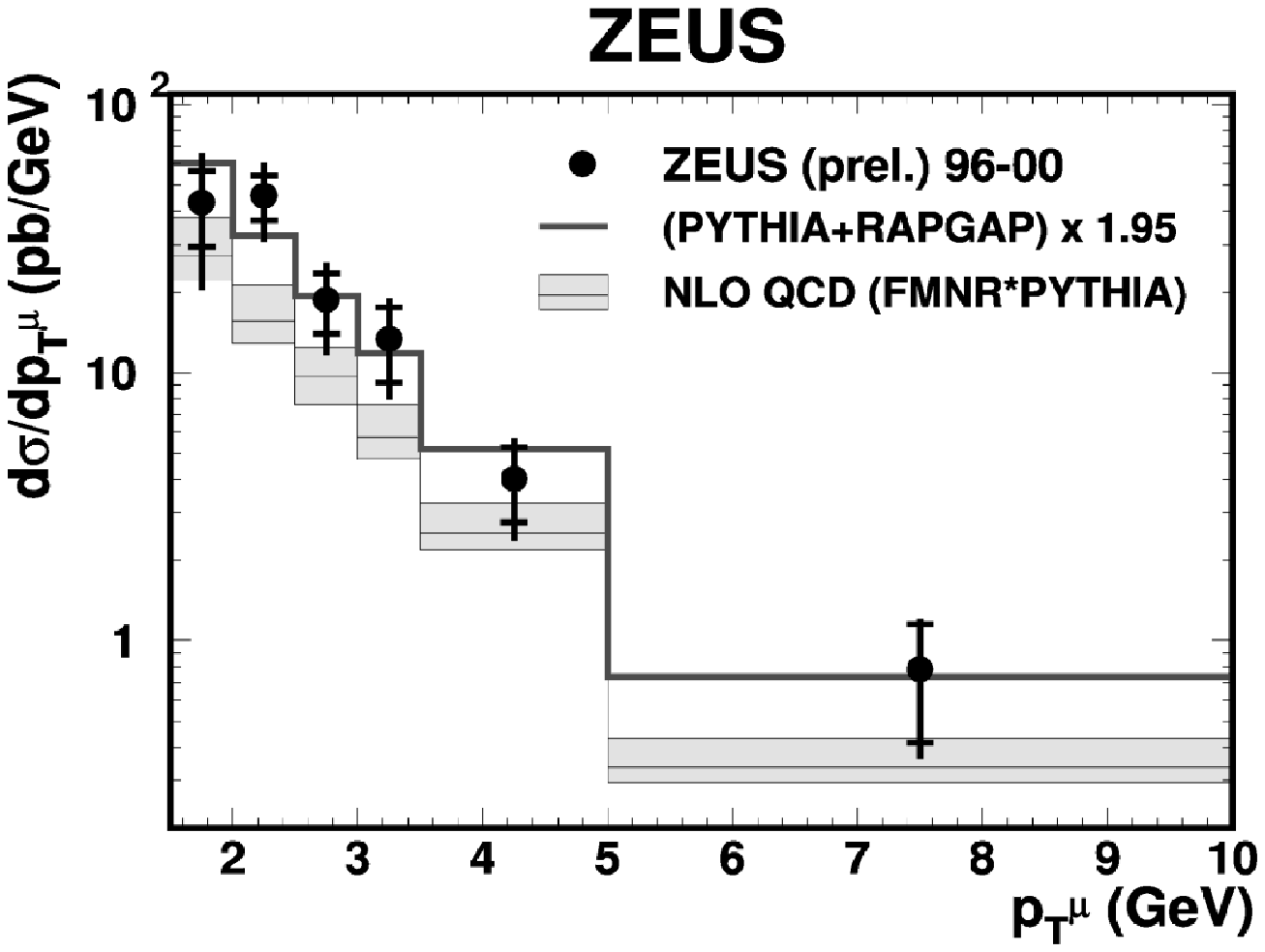}
\includegraphics[width=0.4\textwidth]{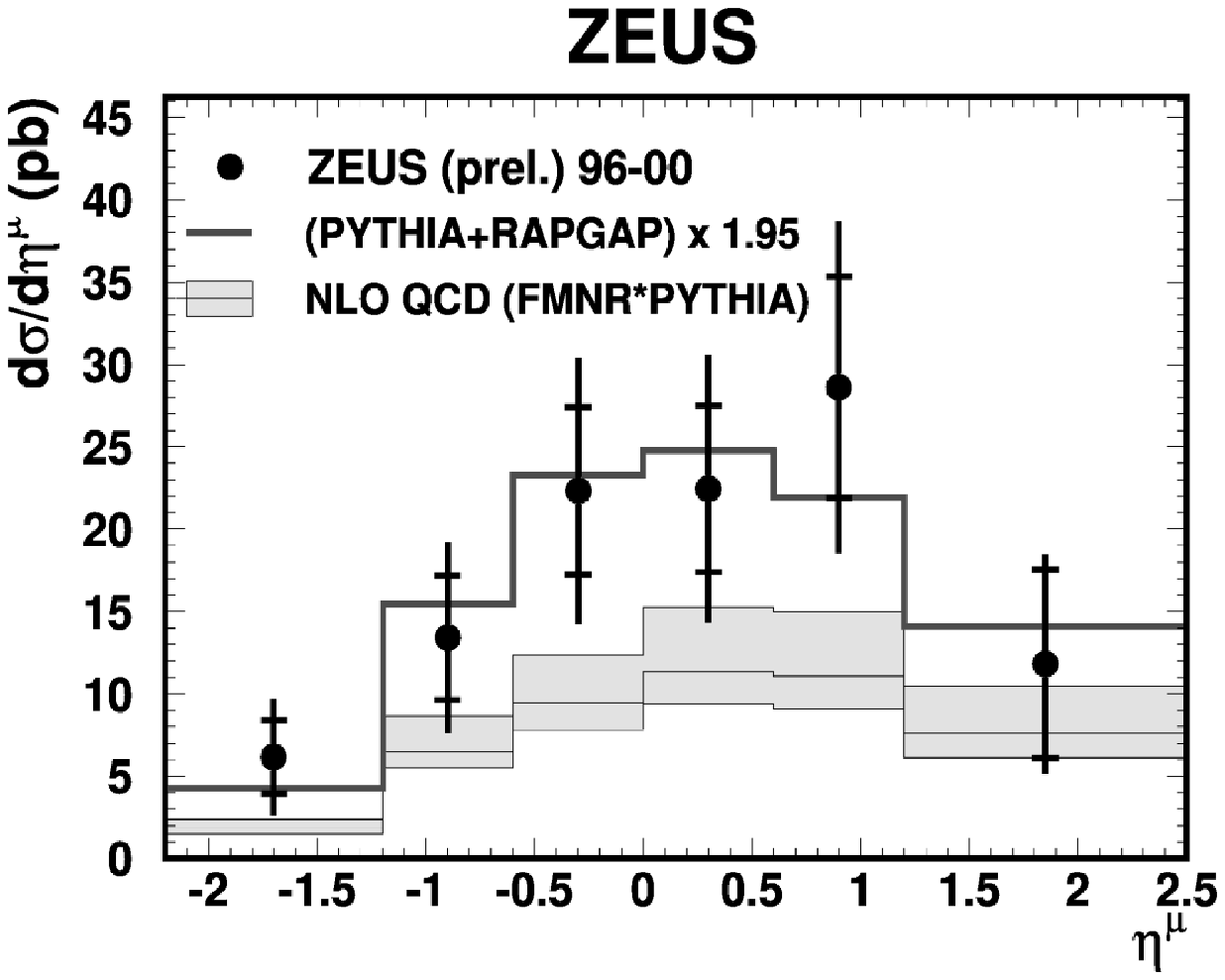}
\caption{Differential cross sections as a 
function of the muon 
transverse momentum $p_{T}^{\mu}$ (on the left) and the muon 
pseudo-rapidity $\eta^{\mu}$ (on the right), for muons from $b$ 
decays in dimuon events. Data come from the ZEUS experiment. 
The full error bars are the 
quadratic sum of the statistical (inner part) and systematic 
uncertainties. The data are compared to the NLO QCD predictions 
(shaded band) and to the MC predictions (histogram).}
\label{fig:Bmumu}
\end{center}
\end{figure}

As said in section \ref{sec:Beauty-tagging}, using double tagged events
\cite{BH1ZEUSdstarmu,BZEUSmumu} 
it is possible to measure the $b$ production up to very low $p_t$ values. 
In Fig. \ref{fig:Bmumu} the differential cross sections as a function 
of the muon transverse momentum $p_{T}^{\mu}$ (plot on the left) and the muon 
pseudo-rapidity $\eta^{\mu}$ (plot on the right), for muons from $b$ 
decays in dimuon events and restricted to the phase space 
$p_{t}^{\mu} >$ 1.5 GeV and -2.2$ < \eta^{\mu} < $ 2.5 for both muons are 
shown. Very good agreement is observed with the PYTHIA+RAPGAP \cite{RAPGAP}  
predictions scaled by a factor 1.95 (histogram). 
Apart from the normalization, the leading 
parton shower approach yields a good description of the corresponding 
physics processes within the entire accessible phase space. 
The data are also compared to the absolute NLO prediction in the massive 
scheme convoluted with the hadronization from PYTHIA MC (shaded band). 
Again, good agreement in shape is observed, with a tendency to 
underestimate the data normalisation. A potential trend for increasing 
data/theory deviations towards low $p_t$ and/or high $\eta$, suggested by 
other measurements as said before, is not supported. 
      
\begin{figure}[htb]
\begin{center}
\includegraphics[width=0.45\textwidth]{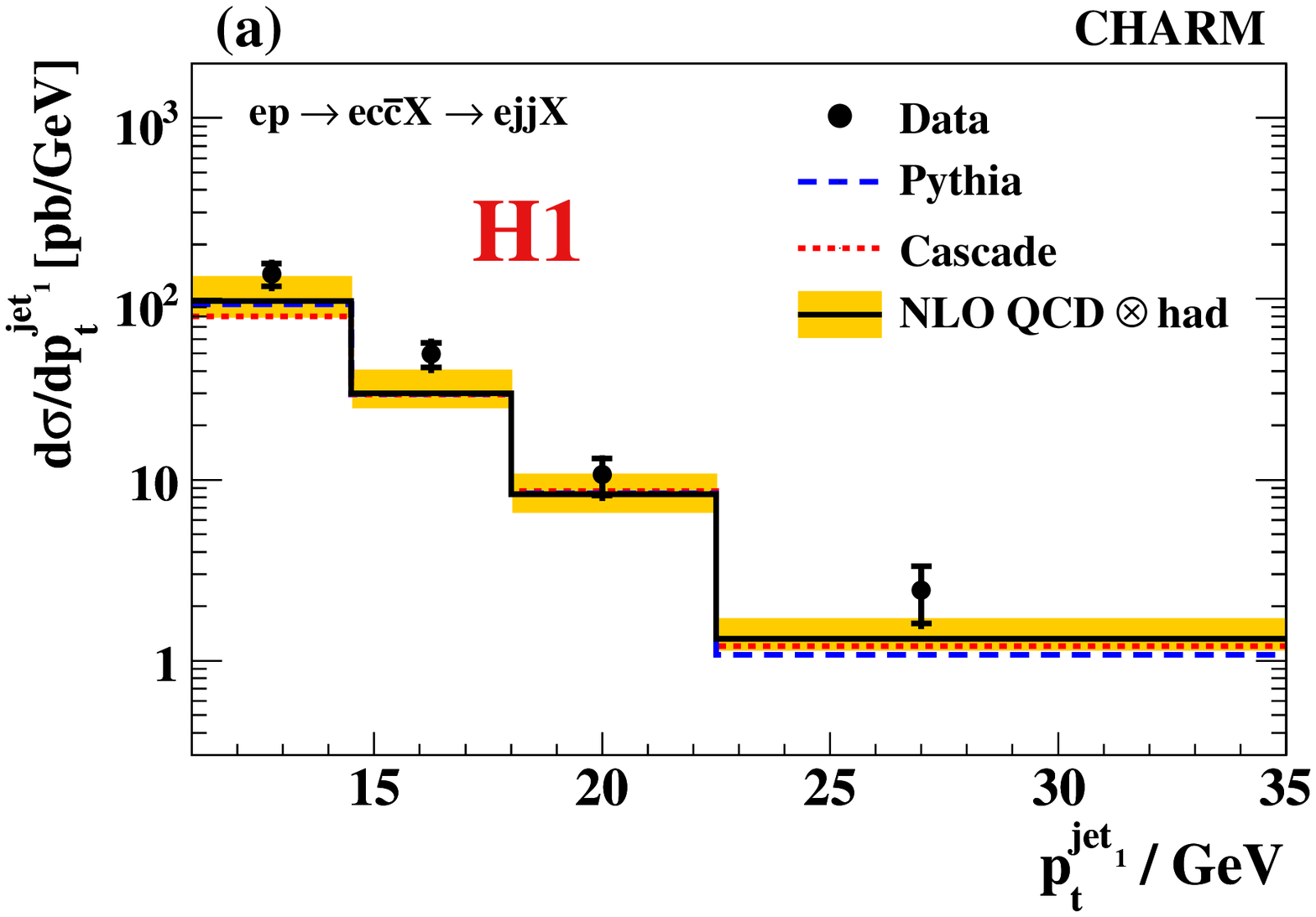}
\includegraphics[width=0.45\textwidth]{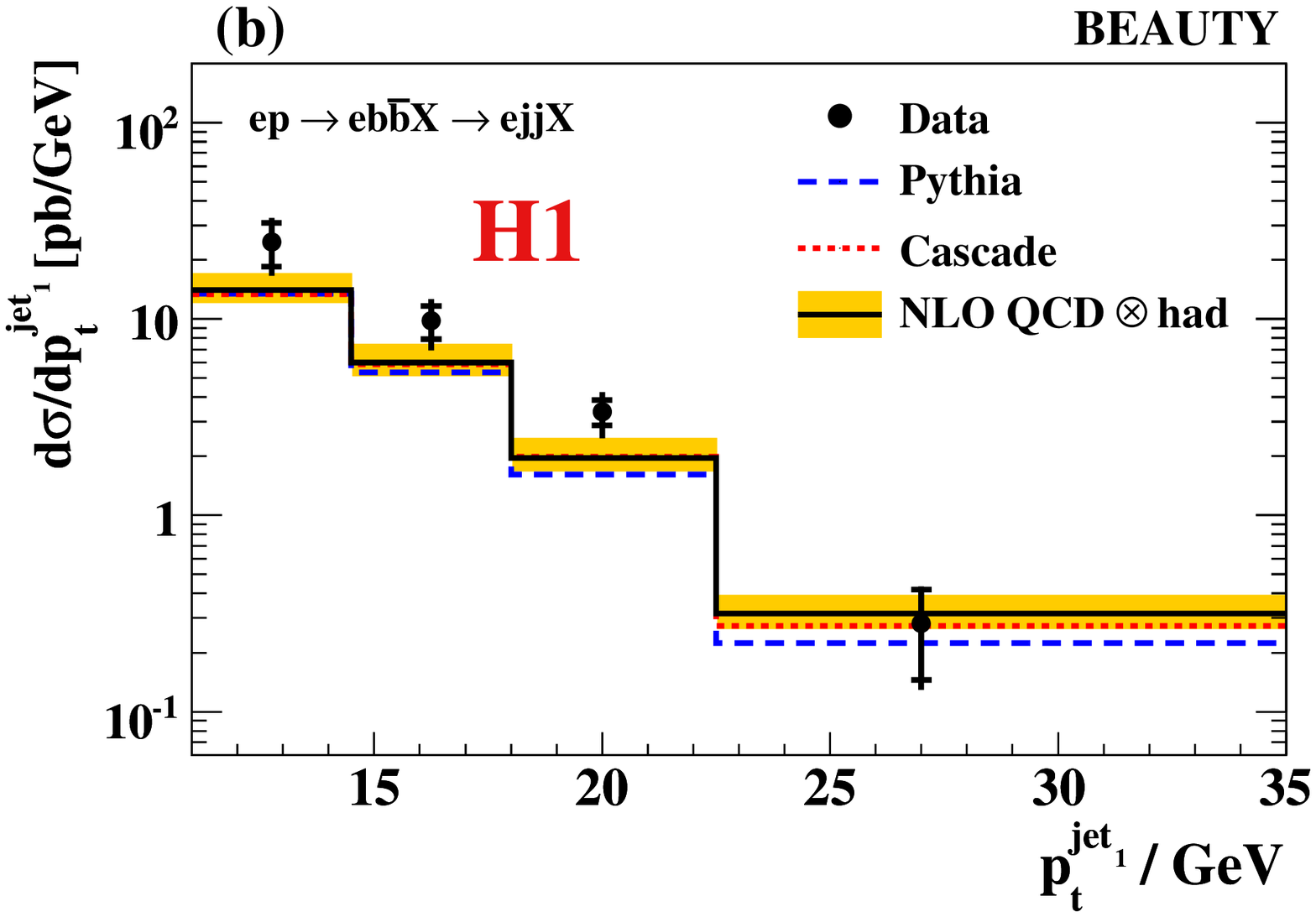}
\end{center}
\caption{Differential charm (on the left) and beauty (on the right) 
photoproduction cross sections $d\sigma/dp_t^{jet_1}$ for the 
process $ep \rightarrow e(c\bar{c}$ or $b\bar{b})X \rightarrow ejjX$. 
The full error bars are the 
quadratic sum of the statistical (inner part) and systematic 
uncertainties. The solid lines indicate the prediction from a NLO 
QCD calculation, corrected for hadronisation effects, and the shaded 
band shows the estimated uncertainty. The absolute predictions from 
PYTHIA (dashed lines) and CASCADE (dotted lines) are also shown.}
\label{fig:Bdijetsvtx}
\end{figure}

Exploiting the experimental possibilities offered by its microvertex 
detector, H1 has measured charm and beauty photoproduction using 
events with two or more jets at high transverse momentum \cite{BH1dijetvtx}. 
In this analysis events containing heavy quarks are distinguished from light 
quark events by the long lifetime of $c$ and $b$ flavoured hadrons, 
which lead to the displacements of tracks from the primary vertex (see 
section \ref{sec:Beauty-tagging}). This analysis provides the first 
simultaneous measurement of charm and beauty in photoproduction, 
extending to larger values of transverse jet momentum than previous 
measurements.  
In Fig. \ref{fig:Bdijetsvtx} the measured differential cross sections for 
charm (plot on the left) and beauty (plot on the right) as functions of the 
transverse momentum of the leading jet $p_{t}^{jet_1}$ are shown. 
Both charm and beauty data are reasonbly well described in shape both by  
the Monte Carlo simulations (PYTHIA and 
CASCADE\footnote{The CASCADE program implements the $k_t$-factorisation approach 
instead of the usual collinear factorisation approach. In the $\gamma g^{*} 
\rightarrow Q \bar{Q} $ matrix element, which takes the heavy quark mass into 
account, the incoming gluon is treated off mass-shell and can have a finite 
transverse momentum. The calculations are performed at LO, higher order QCD 
corrections are simulated with initial state parton showers.} 
\cite{CASCADE}) and the NLO QCD (FMNR) calculations. For charm, the 
NLO QCD calculation is somewhat lower than the measurement but still in 
reasonable agreement within the theoretical errors, for beauty the 
disagreement is slightly higher. The MC's  predict a normalisation which 
is similar to that of FMNR. 
The bulk of the disagreement between data and NLO calculation, 
especially for the beauty, is observed in the region of small values of 
$x_{\gamma}^{obs}$ where the prediction lies below the data. Restricting 
the data to $x_{\gamma}^{obs} >$ 0.85, a significant improvement can be 
obtained: the charm cross sections are in good agreement with the NLO QCD 
calculation both in normalisation and shape, the beauty cross sections are 
also reasonably well described.   
 
\begin{figure}[htb]
\begin{center}
\includegraphics[width=0.6\textwidth]{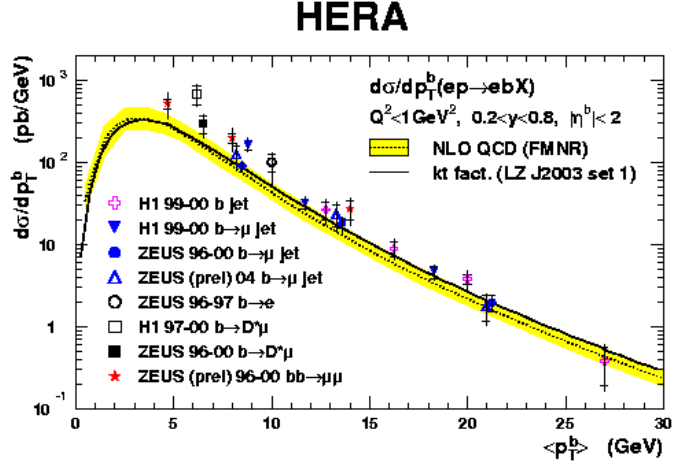}
\end{center}
\caption{Differential cross section for $b$-quark production as a function 
of the $b$-quark transverse momentum $p_{T}^{b}$ for $b$-quark 
pseudo-rapidity $|\eta^{b} | <2$ 
and for $Q^2 < 1$ GeV$^2$, $0.2<y< 0.8$. 
The various points show results from the H1 and ZEUS Collaborations 
using different $b$-tag methods. The full error bars are the 
quadratic sum of the statistical (inner part) and systematic 
uncertainties. The dashed line shows the NLO QCD prediction with the 
theoretical uncertainty shown as the shaded band. The continuous line shows 
the $k_t$ factorization predictions from CASCADE MC.}
\label{fig:Bsummary}
\end{figure}

The major part of the results shown in this section were obtained in the  
photoproduction regime ($Q^2 < 1$ GeV$^2$), and they differ greatly due to 
different experimental cuts, different tagging-methods. It is difficult 
to compare each other and also to extract a general message from the 
comparison between data and NLO QCD calculations. 
In order to overcome these difficulties the various measured cross sections 
were translated to $b$-quark differential cross sections as a function 
of the quark transverse momentum, $d\sigma(ep \rightarrow bX)/dp_{T}^{b}$, 
in the pseudo-rapidity range $|\eta^b| < 2$. 
In Fig. \ref{fig:Bsummary} the so extrapolated differential cross sections 
are shown and compared with the NLO QCD (FMNR) calculations (shaded band). 
The data are in reasonable agreement between them, they 
tend to be somewhat higher than the predictions, the disagreement is  
concentrated at low and medium values of $p_{T}^{b}$, at high values there 
is a nice agreement.

\section{The charm and beauty structure functions}
\label{sec:F2}

The structure functions more frequently studied ($F_2$ and $xF_3$) 
are inclusive objects 
and thus contain contributions from both valence and sea quarks. 
The H1 and ZEUS detectors have the ability to provide identification 
of a particular quark flavour opening so the possibility of 
studying the contribution of that flavour to $F_2$. This is particularly 
important in the case of heavy flavours, as they are likely produced 
in the hard scattering and not in 
the subsequent hadronisation of the struck parton. In other words 
very precise theoretical predictions can be done as explained in 
the section \ref{sec:Theory}. 
Due to the fact that at order $\alpha_s$ heavy quark production in DIS 
occurs through boson-gluon fusion process (see Fig. \ref{fig:feynman}), 
this process involves the gluon density $xg$ directly so it gives an 
experimental handle on this quantity. 

$F_{2}^{c\bar{c}}$ 
is calculated from the measured charm cross sections as follows:
\begin{itemize}
\item
The cross section for $c\bar{c}$ is calculated from the $D^{*}$ cross section 
\cite{F2charmD} (extrapolated to the full phase space) using:
\begin{equation}
\sigma(ep \rightarrow e c\bar{c} X) = \frac{1}{2} \frac{
\sigma(ep \rightarrow e D^{*} X)}{P(c \rightarrow D^{*})} 
\end{equation}
where $P(c \rightarrow D^{*})$ is the probability that a charm quark will 
produce a $D^{*}$ meson (about 25\%). 
As said in the sections \ref{sec:Charm-tagging} and 
\ref{sec:Beauty-tagging}, the advent of the micro-vertex detectors 
has permitted to distinguish events containing heavy quarks from 
light quark events by the long lifetimes of $c$ and $b$ flavoured hadrons, 
which lead to displacements of tracks from the primary vertex. Furthermore  
the results can be obtained in kinematic regions where there is little 
extrapolation needed to the full phase space and so the model 
dependent uncertainty due to the extrapolation is small. These 
measurements were done by the H1 Collaboration \cite{F2charmbeauty}.
 
\item
Finally $F_{2}^{c\bar{c}}$ is related to $ep \rightarrow e c\bar{c}X$ 
cross-section by:
\begin{equation}
\frac{d^2 \sigma(c\bar{c})}{dx dQ^2}= \frac{2\pi\alpha^2}{Q^4x}((1+(1-y)^2)
F_{2}^{c\bar{c}} -y^{2}F_{L}^{c\bar{c}}),
\end{equation}
where the small contribution from $F_{L}^{c\bar{c}}$ is calculated from QCD, 
while $xF_3$ is neglected due to the fact that the measurements are made at 
small $Q^2$.
\end{itemize}
   
\begin{figure}[htb]
\begin{center}
\includegraphics[width=0.49\textwidth]{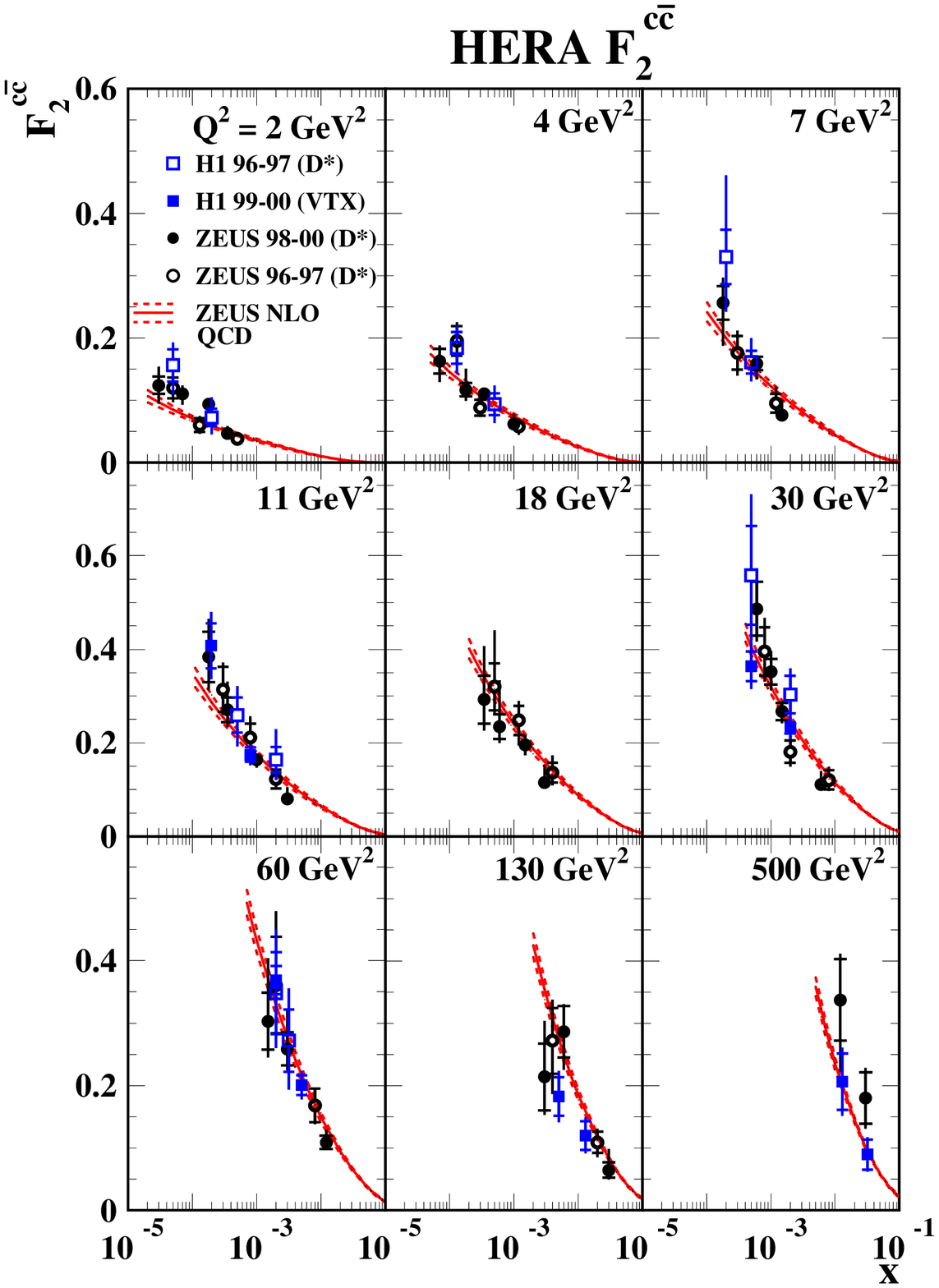}
\includegraphics[width=0.49\textwidth]{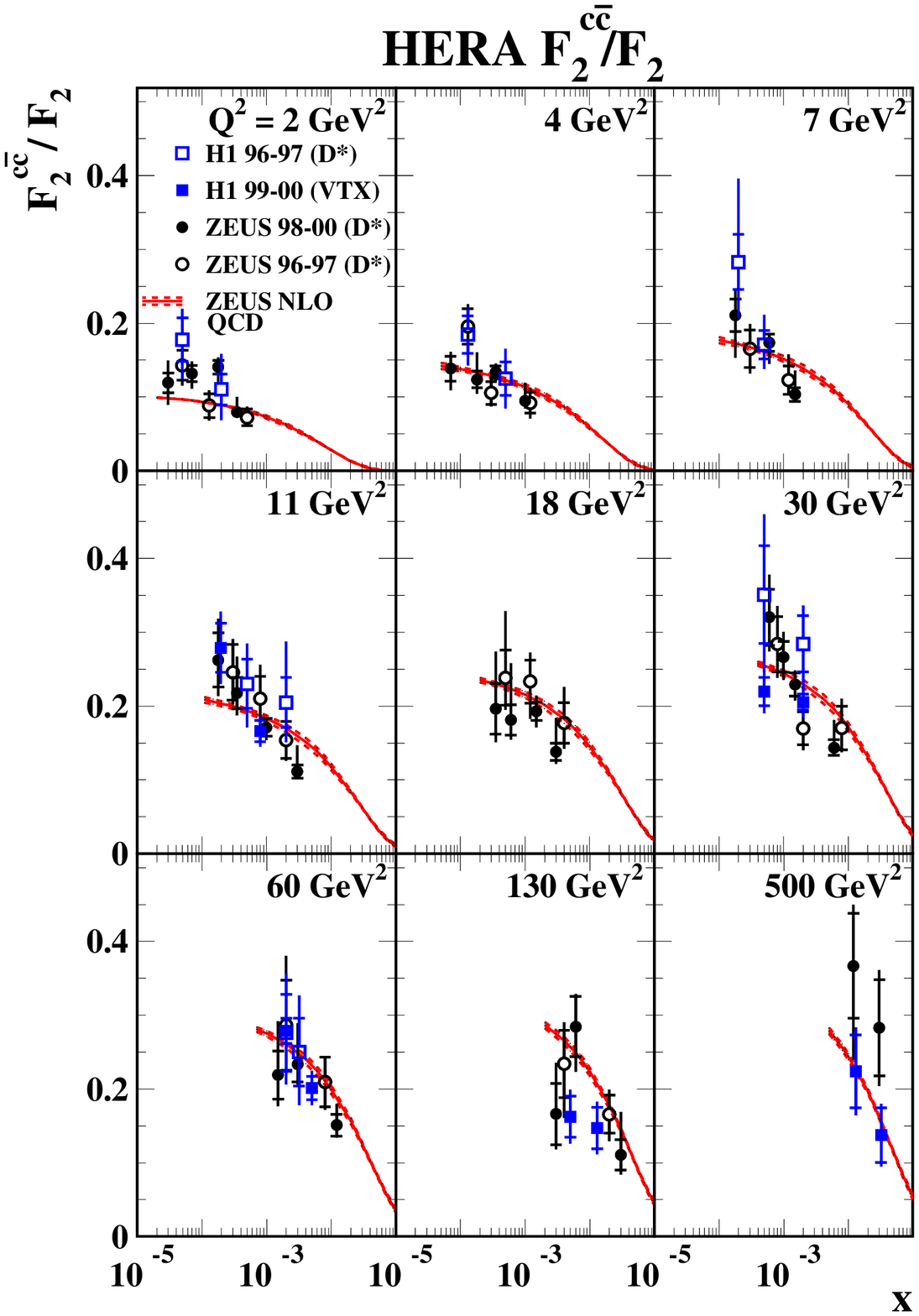}
\end{center}
\caption{On the left plot, the measured $F_2^{c\bar{c}}$ at $Q^2$ values 
between 2 and 500 GeV$^2$ as a function of $x$ is shown while on the right 
plot the measured ratio $F_2^{c\bar{c}}/F_2$ . 
Data from the H1 and ZEUS experiments using different charm tagging are shown.
The data are shown with statistical uncertainties (inner bars) and 
statistical and systematic uncertainties added in quadrature (outer bars). 
The curves represent the ZEUS NLO fit.}
\label{fig:F2charm}
\end{figure}

In Fig. \ref{fig:F2charm} (plot on the left) 
all the data about $F_{2}^{c\bar{c}}$ are shown as function of $x$ at 
$Q^2$ values between 2 and 500 GeV$^2$. The various data sets, obtained with 
different techniques, are in good agreement between them. 
The structure function  $F_{2}^{c\bar{c}}$ shows a rise with 
decreasing $x$ at constant values of $Q^2$. The rise becomes steeper at 
higher $Q^2$.   
The data are compared to calculations using the recent ZEUS NLO fit 
\cite{ZEUSNLO}, in which the parton densities in the proton are 
parameterized by performing fits to inclusive DIS measurements from ZEUS and 
fixed-target experiments. The prediction describes the data well for all 
$Q^2$ and $x$ except for the lowest $Q^2$, where some difference is observed. 
In Fig. \ref{fig:F2charm} (plot on the right) the ratio $F_{2}^{c\bar{c}}/F_2$ 
is shown as function of $x$ at fixed values of $Q^2$. The charm 
contribution to $F_2$ rises from 10\% to 30\% as $Q^2$ increases and $x$ 
decreases. The strong rise of $F_{2}^{c\bar{c}}$ at low values of $x$ is 
similar to that of the gluon density and thus supports the hypothesis that 
charm production is dominated by the boson-gluon fusion mechanism.   
 
\begin{figure}[htb]
\begin{center}
\includegraphics[width=0.5\textwidth]{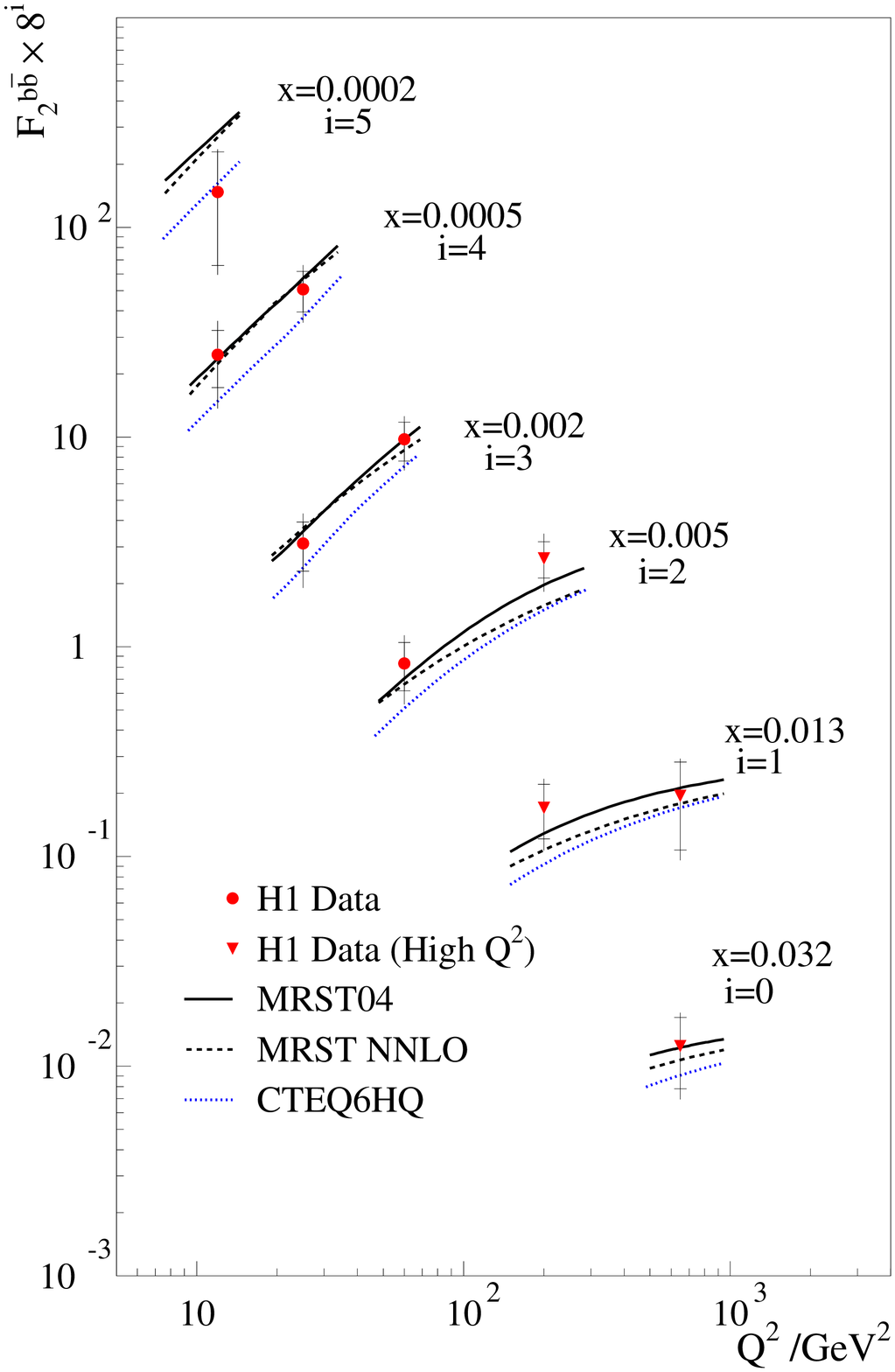}
\end{center}
\caption{The measured $F_{2}^{b\bar{b}}$ shown as function of $Q^2$ for 
various $x$ values. The inner error bars show the statistical errors, 
the outer error bars represent the statistical and systematic errors added in
quadrature. The prediction of QCD are also shown.}
\label{fig:F2beauty}
\end{figure}

Using the help of the micro-vertex detector it was possible 
to measure the structure function $F_2^{b\bar{b }}$ \cite{F2charmbeauty} 
in a similar manner to those depicted for the $F_2^{c\bar{c }}$. 
The measurement of the $b$ cross section (and so of $F_2^{b\bar{b }}$) 
is particularly challenging 
since $b$ events comprise only a small fraction (typically $< 5\%$) of the 
total cross section.
In Fig. \ref{fig:F2beauty} the measured $F_{2}^{b\bar{b}}$ 
(by the H1 Collaboration) is shown as function of $Q^2$. The measurement shows 
positive scaling violations which increase with decreasing of $x$. 
The data are compared with the variable flavour number scheme 
QCD predictions from MRST \cite{MRST} and CTEQ \cite{CTEQ} at NLO and 
a recent calculation at NNLO \cite{NNLO}. The predictions are found to describe 
the data reasonably well. The beauty contribution to $F_2$, in the present 
kinematic range, increases rapidly with $Q^2$ from 0.4\% at $Q^2 = 12$ GeV$^2$ 
to 1.5\% at $Q^2 = 60$ GeV$^2$.

\section{Polarized gluon distribution}
\label{sec:COMPASS}
\begin{figure}[htb]
\begin{center}
\includegraphics[width=0.7\textwidth]{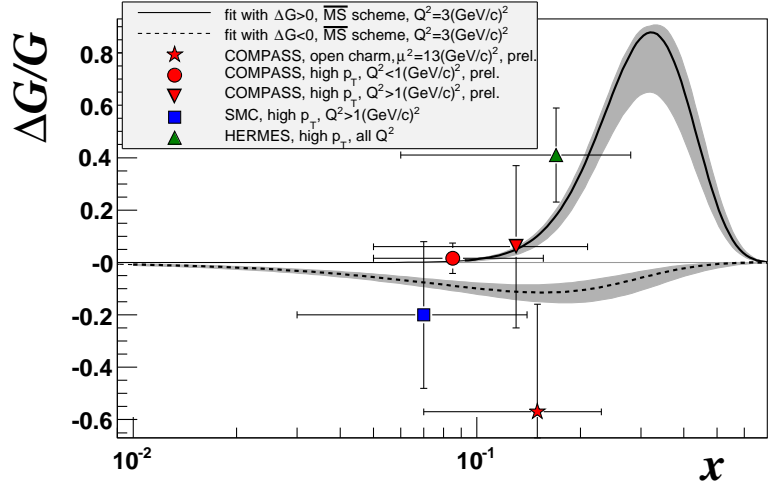}
\end{center}
\caption{Gluon polarization $\Delta g/g$ as a function of $x$ at $Q^2 = Q^{2}_{0}$ 
obtained by NLO QCD fits (bands) and from LO analyses of hadron helicity 
asymmetries (symbols).}
\label{fig:DeltaG}
\end{figure}

In this section, results obtained by the COMPASS Collaboration \cite{COMPASS} 
on the determination 
of the polarized gluon distribution $\Delta g$ using the open charm processes 
in polarized deep inelastic scattering are presented. Formally, one may write for the spin 
of the proton:
\begin{equation}
\frac{1}{2} = \frac{1}{2}\Delta\Sigma + \Delta g + <L_z>,
\end{equation}
where $\Delta\Sigma$ is the contribution from the quarks and antiquarks,  $\Delta g$ 
from the gluons and the last term the mean contribution of any orbital angular 
momentum of the constituents. While for $\Delta\Sigma$ the situation starts to be 
solid, the challenge remains to measure $\Delta g$ and $<L_z>$. 
The COMPASS experiment at CERN is a facility for spectroscopy and spin physics using 
hadron and muon beams from the SpS with a variety of targets and a range of 
sophisticated detectors for analyzing the final state. 
With a wide range of particle identification devices, the 
measurement of $\Delta g$ through the boson-gluon fusion production of $c\bar{c}$ 
pairs is a primary aim. This first measurement was performed by scattering a positive 
muon polarized beam at 160 GeV on a solid polarized target. COMPASS has searched 
for $D^o$ mesons in the decay $D^o \rightarrow K^{-}\pi^{+}$. To reduce background 
the neutral $D$'s was also tagged by requiring them to come from the decay 
$D^{*+} \rightarrow D^o \pi^{+}$. In the measurement there is no reconstruction 
of the $D^o$ vertex, all the reconstruction is based on the 
determination of the invariant mass and 
in the identification of the kaon through the RICH detector. 
The result is $\Delta g/g = -0.57 \pm 0.41$(stat) at a $x$ value of the gluon equal 
to 0.15 and at a $Q^2$ = 13 GeV$^2$; the systematic error is smaller than the statistical 
one.
In Fig. \ref{fig:DeltaG} the gluon polarization $\Delta g/g$ as a function of $x$ at 
fixed $Q^2$ is shown. The points represent the present LO analyses of hadron 
helicity asymmetries (mainly from high $p_T$ hadrons). The result from open charm 
obtained by COMPASS is also shown (star symbol). It is smaller than - but still 
compatible with - zero.       
COMPASS performed a NLO fits to the spin-dependent structure function $g_1(x,Q^2)$ 
world data. Two about equally good solutions for $\Delta g(x,Q^2)$ were found, one 
with a positive and one with a negative first moment $\Delta G$.

\section{Conclusions}
\label{sec:conclusions}
In the previous pages, part of the results obtained 
by the H1 and ZEUS Collaborations in the field of heavy flavours 
has been summarized. 
We have seen that their charm and beauty data are in satisfactory agreement. 
In photo-production regime, beauty and charm data are in general agreement with 
the NLO predictions, even if beauty data are partially slightly higher. 
Charm production gives a large contribution  to the inclusive DIS cross section: 
it was measured with good precision in a large part of phase space, NLO 
QCD calculations describe the data within accuracy. 
The first $F_{2}^{b\bar{b}}$ measurement was also shown. 
All the presented results come from the HERA-I period, much more will come 
using all the statistics from HERA-II period. 
In the polarized DIS field the new preliminary result on $\Delta g/ g$ from 
the COMPASS Collaboration using open charm was shown. 
The measurement, considered 
the most model-independent tool to study gluon-polarisation,  still suffers 
from big statistical uncertainties, they will be highly reduced 
using the large 
amount of data that COMPASS will collect in the near future.

\end{document}